\documentclass[10pt,pra,showpacs]{revtex4}
\usepackage{amsmath}
\usepackage[dvips]{graphicx}
\usepackage{amsfonts}
\usepackage{float}
\usepackage{amssymb}

\begin{document}

\title{Experimental preparation of eight-partite linear and two-diamond
shape cluster states for photonic qumodes}
\author{Xiaolong Su, Yaping Zhao, Shuhong Hao, Xiaojun Jia, Changde Xie, and
Kunchi Peng} \email{kcpeng@sxu.edu.cn} \affiliation{State Key
Laboratory of Quantum Optics and Quantum Optics Devices, Institute
of Opto-Electronics, Shanxi University, Taiyuan, 030006, People's
Republic of China}

\begin{abstract}
The preparation of multipartite entangled states is the prerequisite
for exploring quantum information networks and quantum computation.
In this letter, we present the first experimental demonstration of
eight-partite spatially separated CV entangled states. The initial
resource quantum states are eight squeezed states of light, through
the linearly optical transformation of which two types of the
eight-partite cluster entangled states are prepared, respectively.
The generated eight entangled photonic qumodes are spatially
separated, which provide valuable quantum resources to implement
more complicated quantum information task.
\end{abstract}

 \maketitle

Developing quantum information (QI) science have exhibited unusual
potentiality \cite{Nielsen2000,Brau}. Optical QI based on exploiting
discrete-variable (DV) of single-photon states (photonic qubits) and
continuous-variable (CV) of optical modes (photonic qumodes) plays
important role in QI development. The one-way quantum
computation(QC) based on multipartite cluster
entanglement is initially proposed by Raussendorf and Briegel in the DV model%
\cite{Raussendorf2001}, then it is extended to the CV regime by
Menicucci et al \cite{Menicucci2006}. For one-way QC model the
qubits (qumodes) are initialized in a multipartite cluster entangled
state firstly, then a variety of quantum logical operations can be
achieved only via the single-qubit (qumode) projective measurement
and the classical feedforward of the measured outcomes, in which the
order and choices of measurements are determined by the required
algorithm\cite{Raussendorf2001}. The basic logical operations of
one-way DVQC has been experimentally demonstrated by several groups
\cite{Walther2005,Chen,Gao}.

Parallelly, the theoretical and experimental explorations on one-way
CVQC
were also proceeding continually \cite%
{vanLoockJOSA2007,Tan2009,Gu2009,Miwa2009,Wang2010,Ukai2011,Ukai20112}.
In contrast of the probabilistic generation of photonic qubits in
most cases, CV cluster states are produced in an unconditional
fashion and thus the one-way QC with CV cluster entangled photonic
qumodes can be implemented
deterministically \cite%
{Su2007,Yukawa2008,Tan2008,Miwa2009,Wang2010,Ukai2011,Ukai20112,Matt2011}.
Following the theoretical proposals on one-way CVQC the principally
experimental demonstrations of various one-way QC logical operations
over CVs were achieved by utilizing bipartite and four-partite
cluster entangled photonic qumodes, respectively
\cite{Miwa2009,Wang2010,Ukai2011,Ukai20112}. To develop more
complicated QC larger cluster states with more numbers of entangled
qubits (qumodes) are desired. However, the numbers of spatially
separable entangled qumodes generated by
experiments still stay below four-partites, so far \cite%
{Su2007,Yukawa2008,Tan2008}. In the paper, we present the first
experimental achievement on producing CV eight-partite entangled
states for photonic qumodes. Using eight squeezed states of light to
be the initial resource quantum states and passing through the
linearly optical transformation on a specially designed
beam-splitter network, the eight-partite linear and two-diamond
shape cluster states for photonic qumodes are prepared,
respectively. The entanglement feature among the obtained eight
space-separated photonic qumodes is confirmed by the fully
inseparability criteria of CV multipartite entangled states proposed
by van Loock and Furusawa \cite{Loock2003}.

The cluster state is a type of multipartite quantum entangled
graph states corresponding to some mathematic graphs \cite%
{Menicucci2006,Gu2009,Zhang2006}. The CV cluster quadrature
correlations (so-called nullifiers) can be expressed by
\cite{Gu2009,Zhang2006,Loock2007}

\begin{equation}
(\hat{p}_{a}-\sum_{b\in N_{a}}\hat{x}_{b})\rightarrow 0,\qquad
\forall \quad a\in G
\end{equation}%
where $\hat{x}_{a}=(\hat{a}+\hat{a}^{\dagger })/2$ and $\hat{p}_{a}=(\hat{a}-%
\hat{a}^{\dagger })/2i$ stand for quadrature-amplitude and
quadrature-phase operators of an optical mode $\hat{a}$,
respectively. The subscript a (b) expresses the designated mode
$\hat{a}$ ($\hat{b}$). The modes of $a\in G$ denote the vertices of
the graph $G$, while the modes of $b\in N_{a}$ are the nearest
neighbors of mode $\hat{a}$. For an ideal cluster state the
left-hand side of equation (1) trends to zero, which stands for a
simultaneous zero eigenstate of the quadrature combination
\cite{Gu2009}. The CV cluster quantum entanglements generated by
experiments are deterministic, but also are imperfect, the
entanglement features of which have to be verified and quantified by
the sufficient conditions for the
fully inseparability of multipartite CV entanglement \cite%
{Su2007,Yukawa2008,Matt2011,Tan2008}. There are different
correlation combinations [left-hand side of equation (1)] in a
variety of CV cluster multipartite entangled states, which reflect
the complexity and rich usability of these quantum systems. The
expressions of the nullifiers for different graph states depend on
their graph configurations.

\begin{figure}[tbp]
\centerline{
\includegraphics[width=8cm]{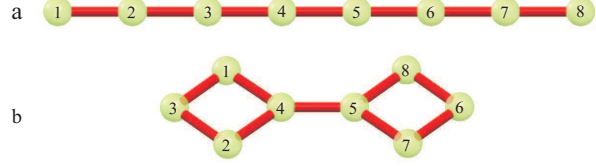}
} \caption{The graph representation of eight-partite cluster states.
a: linear cluster state, b: two-diamond shape cluster state. Each
cluster node corresponds to an optical mode. The connected lines
between neighboring nodes stand for the interaction among these
nodes.}
\end{figure}

Figure 1 (a) and (b) show the graph representations of CV
eight-partite linear (a) and two-diamond (b) shape CV cluster
states, respectively, each node of which corresponds to an optical
mode and the connection lines between neighboring nodes stand for
the interaction between the connected two nodes. From equation (1)
and Fig. 1, we can write the nullifiers of the
linear and the two-diamond shape CV cluster states, respectively, which are $%
\hat{p}_{L1}-\hat{x}_{L2}=\delta _{L1}$, $\hat{p}_{L2}-\hat{x}_{L1}-\hat{x}%
_{L3}=\delta _{L2}$, $\hat{p}_{L3}-\hat{x}_{L2}-\hat{x}_{L4}=\delta _{L3}$, $%
\hat{p}_{L4}-\hat{x}_{L3}-\hat{x}_{L5}=\delta _{L4}$, $\hat{p}_{L5}-\hat{x}%
_{L4}-\hat{x}_{L6}=\delta _{L5}$, $\hat{p}_{L6}-\hat{x}_{L5}-\hat{x}%
_{L7}=\delta _{L6}$, $\hat{p}_{L7}-\hat{x}_{L6}-\hat{x}_{L8}=\delta _{L7}$, $%
\hat{p}_{L8}-\hat{x}_{L7}=\delta _{L8}$ for the linear states; and $\hat{p}%
_{D1}-\hat{x}_{D3}-\hat{x}_{D4}=\delta _{D1}$, $\hat{p}_{D2}-\hat{x}_{D3}-%
\hat{x}_{D4}=\delta _{D2}$,
$\hat{p}_{D3}-\hat{x}_{D1}-\hat{x}_{D2}=\delta
_{D3}$, $\hat{p}_{D4}-\hat{x}_{D1}-\hat{x}_{D2}-\hat{x}_{D5}=\delta _{D4}$, $%
\hat{p}_{D5}-\hat{x}_{D4}-\hat{x}_{D7}-\hat{x}_{D8}=\delta _{D5}$, $\hat{p}%
_{D6}-\hat{x}_{D7}-\hat{x}_{D8}=\delta _{D6}$, $\hat{p}_{D7}-\hat{x}_{D5}-%
\hat{x}_{D6}=\delta _{D7}$,
$\hat{p}_{D8}-\hat{x}_{D5}-\hat{x}_{D6}=\delta
_{D8}$ for the two-diamond states, where the subscripts L$_{i}$ and D$_{i}$ (%
$i=1,2,...8$) denote the individual nodes of the linear and the
two-diamond shape cluster states, respectively, $\delta _{L_{i}}$
and $\delta _{D_{i}}$ express the excess noises resulting from the
imperfect quantum correlations. When the variance of $\delta
_{L_{i}}$ ($\delta _{D_{i}}$) is smaller than the corresponding
quantum noise limit (QNL) determined by vacuum noises, the
correlations among the combined optical modes is within the quantum
region, otherwise the quantum correlations do not exist.

The schemes of generating CV multipartite entangled states commonly
used in experiments are to achieve a linearly optical transformation
of input squeezed states on a specific beam-splitter network
\cite{Loock2007}. Assuming $\hat{a}_{l}$ and $U_{kl}$ stand for the
input squeezed states and the unitary matrix of a given
beam-splitter network, respectively, the
output optical modes after the transformation are given by $\hat{b}%
_{k}=\sum_{l}U_{kl}\hat{a}_{l}$, where the subscripts l and k
express the designated input and output modes, respectively. In our
experiment, four
quadrature-amplitude $\hat{x}$-squeezed states, $\hat{a}_{m}=e^{-r}\hat{x}%
_{m}^{(0)}+ie^{+r}\hat{p}_{m}^{(0)}$ $(m=1,3,5,7)$, and four
quadrature-phase $\hat{p}$-squeezed states, $\hat{a}_{n}=e^{+r}\hat{x}%
_{n}^{(0)}+ie^{-r}\hat{p}_{n}^{(0)}$ $(n=2,4,6,8)$, are applied, where $\hat{%
x}_{j}^{(0)}$\ and $\hat{p}_{j}^{(0)}$ denote the
quadrature-amplitude and the quadrature-phase operators of the
corresponding vacuum field, respectively, $r$ is the squeezing
parameter to quantify the squeezing level, $r=0$ and $r=+\infty $
correspond to the two cases of no squeezing and the ideally perfect
squeezing,respectively. The unitary matrix for generating the CV
eight-partite linear cluster state by combining eight squeezed
states on optical beam splitters equals to (see Supplementary
Material)

\begin{equation}
U_{L}=\left(
\begin{array}{cccccccc}
\frac{i}{\sqrt{2}} & \frac{i}{\sqrt{3}} & \frac{i}{\sqrt{10}} & \sqrt{\frac{3%
}{170}} & \sqrt{\frac{5}{102}} & 0 & 0 & 0 \\
\frac{-1}{\sqrt{2}} & \frac{1}{\sqrt{3}} & \frac{1}{\sqrt{10}} & -i\sqrt{%
\frac{3}{170}} & -i\sqrt{\frac{5}{102}} & 0 & 0 & 0 \\
0 & \frac{i}{\sqrt{3}} & -i\sqrt{\frac{2}{5}} & -\sqrt{\frac{6}{85}} & -%
\sqrt{\frac{10}{51}} & 0 & 0 & 0 \\
0 & 0 & \sqrt{\frac{2}{5}} & 3i\sqrt{\frac{3}{170}} &
i\sqrt{\frac{15}{34}}
& 0 & 0 & 0 \\
0 & 0 & 0 & \sqrt{\frac{15}{34}} & -3\sqrt{\frac{3}{170}} & i\sqrt{\frac{2}{5%
}} & 0 & 0 \\
0 & 0 & 0 & i\sqrt{\frac{10}{51}} & -i\sqrt{\frac{6}{85}} & \sqrt{\frac{2}{5}%
} & \frac{1}{\sqrt{3}} & 0 \\
0 & 0 & 0 & -\sqrt{\frac{5}{102}} & \sqrt{\frac{3}{170}} & \frac{i}{\sqrt{10}%
} & \frac{-i}{\sqrt{3}} & \frac{-i}{\sqrt{2}} \\
0 & 0 & 0 & -i\sqrt{\frac{5}{102}} & i\sqrt{\frac{3}{170}} & \frac{-1}{\sqrt{%
10}} & \frac{1}{\sqrt{3}} & \frac{-1}{\sqrt{2}}%
\end{array}%
\right)
\end{equation}

The unitary matrix in equation (2) expresses an optical
transformation on a beam-splitter network consisting of seven beam
splitters and can be decomposed into
$U_{L}=F_{8}I_{7}(-1)F_{6}^{\dagger }F_{4}I_{3}(-1)F_{2}^{\dagger
}B_{78}^{-}(1/2)F_{8}B_{12}^{-}(1/2)F_{1}B_{67}^{-}(1/3)
F_{7}B_{23}^{-}(1/3)F_{2}B_{56}^{-}(2/5)F_{6}B_{34}^{-}(2/5)F_{3}\allowbreak
B_{45}^{+}(25/34) $, where $F_{k}$ denotes the Fourier
transformation of mode $k$, which corresponds to a 90$^{\circ }$\
rotation in the phase space; $B_{kl}^{\pm }(T_{j})$ stands for the
linearly optical transformation on the jth beam-splitter with the
transmission of $T_{j}$ ($j=1,2,3\ldots ..7$),
where $(B_{kl}^{\pm })_{kk}=\sqrt{1-T}$, $(B_{kl}^{\pm })_{kl}=\sqrt{T}$, $%
(B_{kl}^{\pm })_{lk}=\pm \sqrt{T}$, and $(B_{kl}^{\pm })_{ll}=\mp \sqrt{1-T}%
, $ are elements of beam-splitter matrix. $I_{k}(-1)=e^{i\pi }$
corresponds to a 180$^{\circ }$ rotation of mode $k$ in the phase
space.

\begin{figure}[tbph]
\begin{center}
\includegraphics[width=8cm]{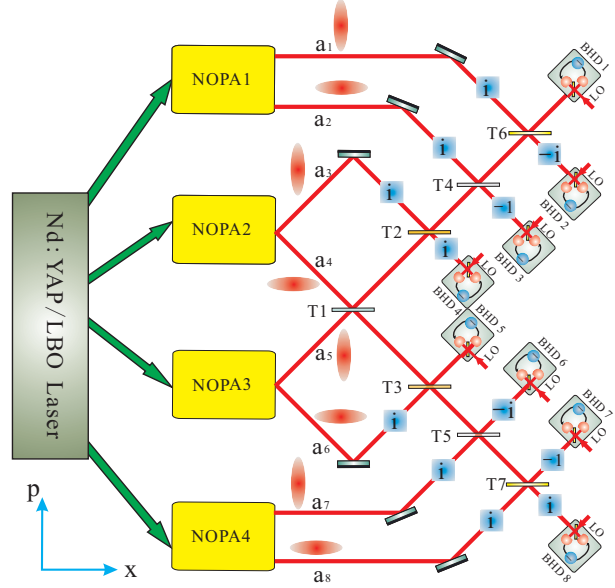}
\end{center}
\caption{Schematic of experimental setup for CV eight-partite
cluster state generation. $T$: transmission efficient of beam
splitter, Boxes including $i$
are Fourier transforms ($90^{\circ}$ rotations in phase space), $-i$ is a $%
-90^{\circ}$ rotation, and $-1$ is a $180^{\circ}$ rotation, BHD:
balanced homodyne detector.}
\end{figure}

Figure 2 shows the schematic of the experimental set-up for
preparing the
eight-partite CV linear cluster state. The four $\hat{x}$-squeezed and four $%
\hat{p}$-squeezed states are produced by four NOPAs pumped by a
common laser source, which is a CW intracavity frequency-doubled and
frequency-stabilized Nd:YAP/LBO(Nd-doped YAlO$_{3}$
perorskite/lithium triborate) with both
outputs of the fundamental and the second-harmonic waves \cite{WangIEEE2010}%
. The output fundamental wave at 1080 nm wavelength is used for the
injected signals of NOPAs and the local oscillators of the balanced
homodyne detectors (BHDs), which are applied to measure the quantum
fluctuations of the quadrature-amplitude and the quadrature-phase of
the output optical modes \cite{Su2007}. The second-harmonic wave at
540 nm wavelength serves as the pump field of the four NOPAs, in
which through an intracavity frequency-down-conversion process a
pair of signal and idler modes with the identical frequency at 1080
nm and the orthogonal polarizations are generated
\cite{Li2002,Wang20102}. Since the amplitude and the phase
quadratures of the signal and the idler modes are entangled each
other, the two coupled modes of them at $\pm 45^{\circ }$
polarization directions both are the squeezed states
\cite{Su2007,Yun2000}. In our experiment, the four NOPAs are
operated at the parametric deamplification situation, i.e. the
phase difference between the pump field and the injected signal is $%
(2n+1)\pi $ ($n$ is an integer). Under this condition, the coupled
modes at +45$^{\circ}$ and -45$^{\circ}$polarization directions are
the quadrature-amplitude and the quadrature-phase squeezed states,
respectively \cite{Su2007,Yun2000}. When the transmissions of the
seven beam splitters are chosen as $T1=25/34$, $T2=T3=2/5$,
$T4=T5=1/3$, $T6=T7=1/2$, the eight output optical modes
$\hat{b_{j}}$ $(j=1,2,...8)$ are in a eight-partite CV
linear cluster state. The quadrature-amplitude and quadrature-phase of each $%
\hat{b_{j}}$ are measured by eight BHDs, respectively. The
nullifiers of the eight output modes depend on the squeezing
parameters of the resource squeezed states. For our experimental
system all four NOPAs have the identical configuration (the
construction of NOPA is described in the Supplementary Material) and
are operated under the same conditions. Each of NOPAs is also
adjusted to produce two balanced squeezed states. So, the eight
initial squeezed states own the same squeezed parameter $r$. In this
case we can easily calculate the excess noises of the nullifiers for
the
eight-partite linear CV cluster state consisting of the eight output modes $%
\hat{b_{j}}$\ ($j=1,.....8$), which are $\delta _{L1}=\sqrt{2}e^{-r}\hat{x}%
_{1}^{(0)}$, $\delta _{L2}=\sqrt{3}e^{-r}\hat{p}_{2}^{(0)}$, $\delta _{L3}=%
\frac{1}{\sqrt{2}}e^{-r}\hat{x}_{1}^{(0)}-\sqrt{\frac{5}{2}}e^{-r}\hat{x}%
_{3}^{(0)}$, $\delta _{L4}=\frac{1}{\sqrt{3}}e^{-r}\hat{p}_{2}^{(0)}+\sqrt{%
\frac{2}{5}}e^{-r}\hat{p}_{6}^{(0)}+\sqrt{\frac{34}{15}}e^{-r}\hat{x}%
_{5}^{(0)}$, $\delta _{L5}=\sqrt{\frac{34}{15}}e^{-r}\hat{p}_{4}^{(0)}-\sqrt{%
\frac{2}{5}}e^{-r}\hat{x}_{3}^{(0)}-\frac{1}{\sqrt{3}}e^{-r}\hat{x}%
_{7}^{(0)} $, $\delta _{L6}=\sqrt{\frac{5}{2}}e^{-r}\hat{p}_{6}^{(0)}-\frac{1%
}{\sqrt{2}}e^{-r}\hat{p}_{8}^{(0)}$, $\delta _{L7}=-\sqrt{3}e^{-r}\hat{x}%
_{7}^{(0)}$ and $\delta _{L8}=-\sqrt{2}e^{-r}\hat{p}_{8}^{(0)}$,
respectively.

The unitary matrix of the two-diamond cluster state $U_{D}$ equals to $%
U_{F}U_{L}$, with $U_{F}=diag\{-1,-i,i,1,1,i,-i,-1\}$ (see
Supplementary Material), thus the two-diamond shape cluster state
can be prepared from the linear cluster state via local Fourier
transforms and phase rotations. The excess noise terms of the
nullifiers of the two-diamond shape cluster
state are expressed by $\delta _{D1}=-\frac{1}{\sqrt{2}}e^{-r}\hat{x}%
_{1}^{(0)}+\sqrt{\frac{5}{2}}e^{-r}\hat{x}_{3}^{(0)}$, $\delta _{D2}=\frac{1%
}{\sqrt{2}}e^{-r}\hat{x}_{1}^{(0)}-\sqrt{\frac{5}{2}}e^{-r}\hat{x}_{3}^{(0)}$%
, $\delta _{D3}=-\sqrt{3}e^{-r}\hat{p}_{2}^{(0)}$, $\delta _{D4}=-\frac{2}{%
\sqrt{3}}e^{-r}\hat{p}_{2}^{(0)}+\sqrt{\frac{2}{5}}e^{-r}\hat{p}_{6}^{(0)}+%
\sqrt{\frac{34}{15}}e^{-r}\hat{x}_{5}^{(0)}$, $\delta _{D5}=\sqrt{\frac{34}{%
15}}e^{-r}\hat{p}_{4}^{(0)}-\sqrt{\frac{2}{5}}e^{-r}\hat{x}_{3}^{(0)}+\frac{2%
}{\sqrt{3}}e^{-r}\hat{x}_{7}^{(0)}$, $\delta _{D6}=\sqrt{3}e^{-r}\hat{x}%
_{7}^{(0)}$, $\delta _{D7}=\sqrt{\frac{5}{2}}e^{-r}\hat{p}_{6}^{(0)}-\frac{1%
}{\sqrt{2}}e^{-r}\hat{p}_{8}^{(0)}$, and $\delta _{D8}=\sqrt{\frac{5}{2}}%
e^{-r}\hat{p}_{6}^{(0)}+\frac{1}{\sqrt{2}}e^{-r}\hat{p}_{8}^{(0)}$,
respectively. According to the inseparability criteria for CV
multipartite entangled states proposed by van Loock and Furusawa
\cite{Loock2003}, we deduced the inseparability criterion
inequalities for CV eight-partite linear and two-diamond shape
cluster states, which are given by equations (3a)-(3g) and equations
(4a)-(4i), respectively (see Supplementary Material).
\begin{widetext}
\begin{subequations}
\begin{eqnarray}
V(\hat{p}_{L1}-\hat{x}_{L2})+V(\hat{p}_{L2}-\hat{x}_{L1}-g_{L3}\hat{x}_{L3})
&<&1 \\
V(\hat{p}_{L2}-g_{L1}\hat{x}_{L1}-\hat{x}_{L3})+V(\hat{p}_{L3}-\hat{x}%
_{L2}-g_{L4}\hat{x}_{L4}) &<&1 \\
V(\hat{p}_{L3}-g_{L2}\hat{x}_{L2}-\hat{x}_{L4})+V(\hat{p}_{L4}-\hat{x}%
_{L3}-g_{L5}\hat{x}_{L5}) &<&1 \\
V(\hat{p}_{L4}-g_{L3}\hat{x}_{L3}-\hat{x}_{L5})+V(\hat{p}_{L5}-\hat{x}%
_{L4}-g_{L6}\hat{x}_{L6}) &<&1 \\
V(\hat{p}_{L5}-g_{L4}\hat{x}_{L4}-\hat{x}_{L6})+V(\hat{p}_{L6}-\hat{x}%
_{L5}-g_{L7}\hat{x}_{L7}) &<&1 \\
V(\hat{p}_{L6}-g_{L5}\hat{x}_{L5}-\hat{x}_{L7})+V(\hat{p}_{L7}-\hat{x}%
_{L6}-g_{L8}\hat{x}_{L8}) &<&1 \\
V(\hat{p}_{L7}-g_{L6}\hat{x}_{L6}-\hat{x}_{L8})+V(\hat{p}_{L8}-\hat{x}_{L7})
&<&1
\end{eqnarray}%
\end{subequations}
and
\begin{subequations}
\begin{eqnarray}
V(\hat{p}_{D1}-\hat{x}_{D3}-g_{D1}\hat{x}_{D4})+V(\hat{p}_{D3}-\hat{x}%
_{D1}-g_{D2}\hat{x}_{D2}) &<&1 \\
V(\hat{p}_{D2}-\hat{x}_{D3}-g_{D1}\hat{x}_{D4})+V(\hat{p}_{D3}-\hat{x}%
_{D2}-g_{D2}\hat{x}_{D1}) &<&1 \\
V(\hat{p}_{D1}-g_{D3}\hat{x}_{D3}-\hat{x}_{D4})+V(\hat{p}_{D4}-\hat{x}%
_{D1}-g_{D4}\hat{x}_{D2}-g_{D5}\hat{x}_{D5}) &<&1 \\
V(\hat{p}_{D2}-g_{D3}\hat{x}_{D3}-\hat{x}_{D4})+V(\hat{p}_{D4}-g_{D4}\hat{x}%
_{1}-\hat{x}_{D2}-g_{D5}\hat{x}_{D5}) &<&1 \\
V(\hat{p}_{D4}-g_{D6}\hat{x}_{D1}-g_{D6}\hat{x}_{D2}-\hat{x}_{D5})+V(\hat{p}%
_{D5}-\hat{x}_{D4}-g_{D6}\hat{x}_{D7}-g_{D6}\hat{x}_{D8}) &<&1 \\
V(\hat{p}_{D5}-g_{D5}\hat{x}_{D4}-\hat{x}_{D7}-g_{D4}\hat{x}_{D8})+V(\hat{p}%
_{D7}-\hat{x}_{D5}-g_{D3}\hat{x}_{D6}) &<&1 \\
V(\hat{p}_{D5}-g_{D5}\hat{x}_{D4}-g_{D4}\hat{x}_{D7}-\hat{x}_{D8})+V(\hat{p}%
_{D8}-\hat{x}_{D5}-g_{D3}\hat{x}_{D6}) &<&1 \\
V(\hat{p}_{D6}-\hat{x}_{D7}-g_{D2}\hat{x}_{D8})+V(\hat{p}_{D7}-g_{D1}\hat{x}%
_{D5}-\hat{x}_{D6}) &<&1 \\
V(\hat{p}_{D6}-g_{D2}\hat{x}_{D7}-\hat{x}_{D8})+V(\hat{p}_{D8}-g_{D1}\hat{x}%
_{D5}-\hat{x}_{D6}) &<&1
\end{eqnarray}%
\end{subequations}
\end{widetext}
where left-hand sides and right-hand sides of these inequalities are
the combination of variances of nullifiers and the boundary,
respectively. In the Supplementary Material, we numerically
calculated the dependencies of the combinations of the correlation
variances in the left-hand sides of
equations (3) and equations (4) on the squeezing factor $r$\ for $g=1$ and $%
g=g^{opt}$ ($g^{opt}$ is the optimal gain factor), respectively. It
can be seen, when the optimal gain factors are used the correlation
variance combinations are always smaller than the boundary for all
values of $r>0$, i.e. the eight-partite CV cluster entanglement
always can be realized by the presented system whatever how low the
squeezing of the initial squeezed states is. However, when $g=1$, a
lower limitation of $r$\ is required to result in the correlation
variances to be smaller than the boundary.

The experimentally measured initial squeezing degrees of the output
fields from four NOPAs are $4.30\pm 0.07$ dB below the QNL which
corresponds to the squeezing parameter $r=0.50\pm 0.02$ (see
Supplementary Material). During the measurements the pump power of
NOPAs at $540$ nm wavelength is $\sim 180$ mW, which is below the
oscillation threshold of $240 $ mW, and the intensity of the
injected signal at $1080$ nm is $10$ mW. The phase difference on
each beam-splitters are locked according to the requirements. The
light intensity of the local oscillator in all BHDs is set to around
5 mW. The measured QNL is about 20 dB above the electronics noise
level, which guarantees that the results of the homodyne detections
are almost not affected by electronic noises.

\begin{figure}[tbph]
\centerline{
\includegraphics[width=10cm,height=7cm]{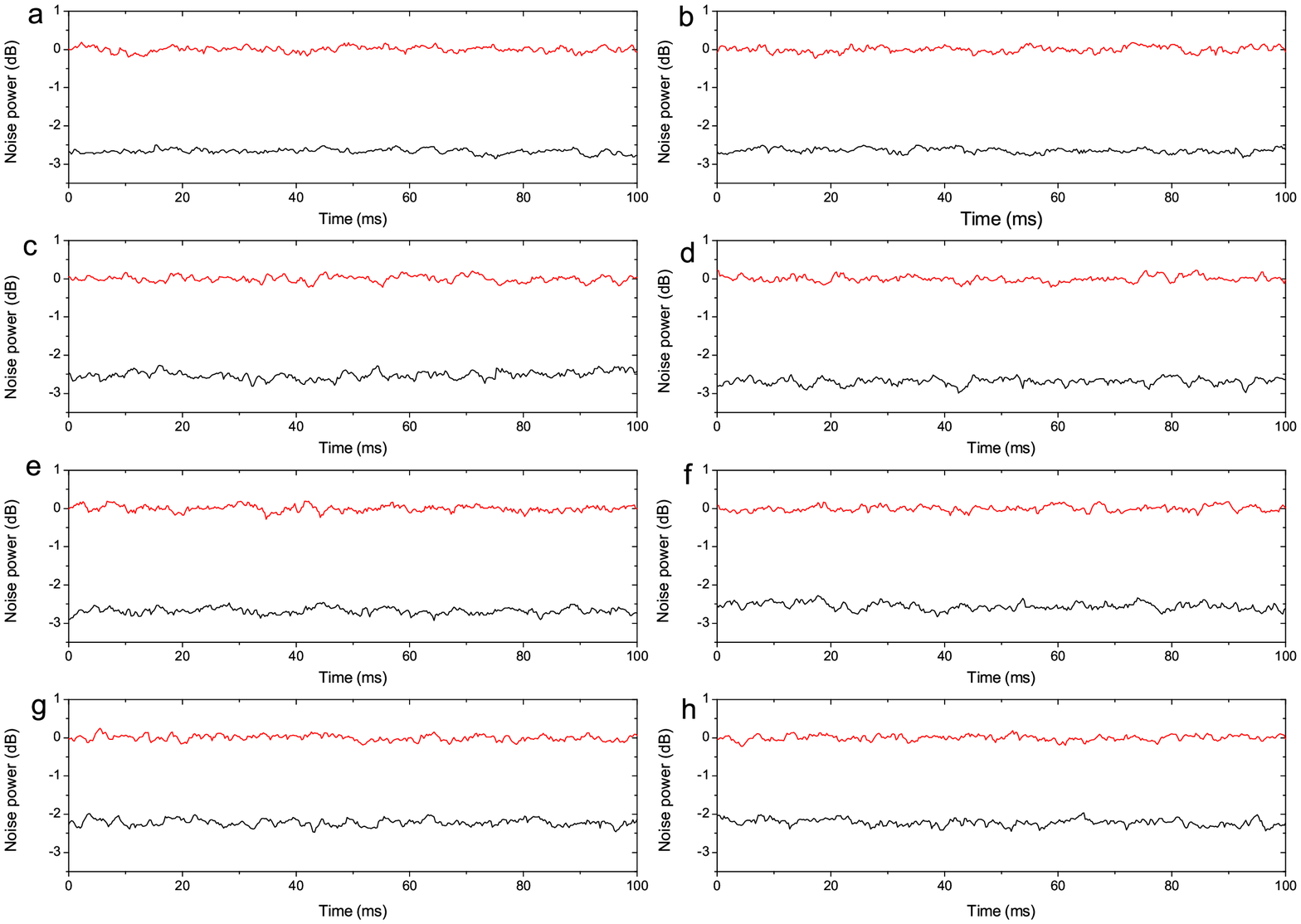}
} \caption{The measured noise powers of eight-partite linear cluster
state. The upper and lower lines in all graphs are shot noise level
and correlation
variances of nullifiers, respectively. (a)-(h) are noise powers of $V(\hat{p}%
_{L1}-\hat{x}_{L2})$, $V(\hat{p}_{L2}-\hat{x}_{L1}-\hat{x}_{L3}) $, $V(\hat{p%
}_{L3}-\hat{x}_{L2}-\hat{x}_{L4})$, $V(\hat{p}_{L4}-\hat{x}_{L3}-\hat{x}%
_{L5}) $, $V(\hat{p}_{L5}-\hat{x}_{L4}-\hat{x}_{L6})$, $V(\hat{p}_{L6}-\hat{x%
}_{L5}-\hat{x}_{L7})$, $V(\hat{p}_{L7}-\hat{x}_{L6}-\hat{x}_{L8}) $, and $V(%
\hat{p}_{L8}-\hat{x}_{L7})$, respectively. The measurement frequency
is 2 MHz, resolution bandwidth is 30 kHz, and video bandwidth is 100
Hz.}
\end{figure}

\begin{figure}[tbph]
\centerline{
\includegraphics[width=10cm,height=7cm]{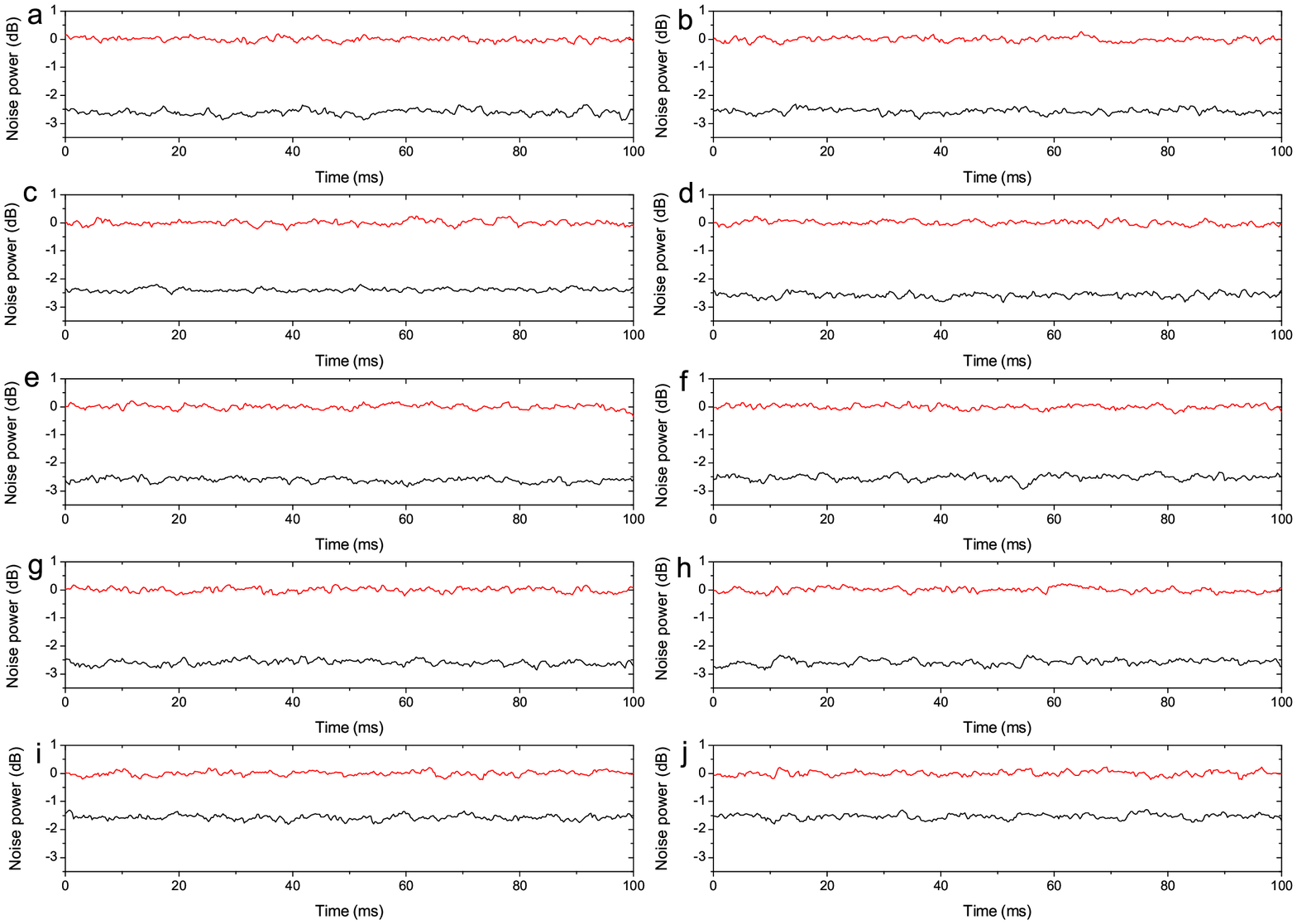}
} \caption{The measured noise powers of eight-partite two-diamond
shape cluster state. The upper and lower lines in all graphs are
shot noise level and correlation variances of nullifiers,
respectively. (a)-(j) are noise
powers of $V(\hat{p}_{D1}-\hat{x}_{D3}-\hat{x}_{D4})$, $V(\hat{p}_{D2}-\hat{x%
}_{D3}-\hat{x}_{D4})$,$\ V(\hat{p}_{D3}-\hat{x}_{D1}-\hat{x}_{D2})$, $V(\hat{%
p}_{D4}-\hat{x}_{D1}-\hat{x}_{D2}-\hat{x}_{D5})$, $V(\hat{p}_{D5}-\hat{x}%
_{D4}-\hat{x}_{D7}-\hat{x}_{D8})$, $V(\hat{p}_{D6}-\hat{x}_{D7}-\hat{x}%
_{D8}) $, $V(\hat{p}_{D7}-\hat{x}_{D5}-\hat{x}_{D6})$, $V(\hat{p}_{D8}-\hat{x%
}_{D5}-\hat{x}_{D6})$, $V(\hat{p}_{D4}-g_{D6}\hat{x}_{D1}-g_{D6}\hat{x}_{D2}-%
\hat{x}_{D5})$, $V(\hat{p}_{D5}-\hat{x}_{D4}-g_{D6}\hat{x}_{D7}-g_{D6}\hat{x}%
_{D8})$, respectively. The measurement frequency is 2 MHz,
resolution bandwidth is 30 kHz, and video bandwidth is 100 Hz.}
\end{figure}

The correlation variances measured experimentally are shown in
figure 3 for
the linear cluster and figure 4 for the two-diamond cluster. They are $V(%
\hat{p}_{L1}-\hat{x}_{L2})=$ $-2.67\pm 0.06$\ dB, $V(\hat{p}_{L2}-\hat{x}%
_{L1}-\hat{x}_{L3})=-2.65\pm 0.13$ dB, $V(\hat{p}_{L3}-\hat{x}_{L2}-\hat{x}%
_{L4})=-2.52\pm 0.20$ dB, $V(\hat{p}_{L4}-\hat{x}_{L3}-\hat{x}%
_{L5})=-2.69\pm 0.09$ dB, $V(\hat{p}_{L5}-\hat{x}_{L4}-\hat{x}%
_{L6})=-2.68\pm 0.08$ dB, $V(\hat{p}_{L6}-\hat{x}_{L5}-\hat{x}%
_{L7})=-2.56\pm 0.10$ dB, $V(\hat{p}_{L7}-\hat{x}_{L6}-\hat{x}%
_{L8})=-2.22\pm 0.09$ dB, $V(\hat{p}_{L8}-\hat{x}_{L7})=-2.21\pm
0.09$ dB
and $V(\hat{p}_{D1}-\hat{x}_{D3}-\hat{x}_{D4})=-2.61\pm 0.10$ dB, $V(\hat{p}%
_{D2}-\hat{x}_{D3}-\hat{x}_{D4})=-2.57\pm 0.09$ dB,$\ V(\hat{p}_{D3}-\hat{x}%
_{D1}-\hat{x}_{D2})=-2.39\pm 0.06$ dB, $V(\hat{p}_{D4}-\hat{x}_{D1}-\hat{x}%
_{D2}-\hat{x}_{D5})=-2.58\pm 0.09$ dB, $V(\hat{p}_{D5}-\hat{x}_{D4}-\hat{x}%
_{D7}-\hat{x}_{D8})=-2.61\pm 0.09$ dB, $V(\hat{p}_{D6}-\hat{x}_{D7}-\hat{x}%
_{D8})=-2.52\pm 0.10$ dB, $V(\hat{p}_{D7}-\hat{x}_{D5}-\hat{x}%
_{D6})=-2.59\pm 0.09$ dB, $V(\hat{p}_{D8}-\hat{x}_{D5}-\hat{x}%
_{D6})=-2.58\pm 0.10$ dB, $V(\hat{p}_{D4}-g_{D6}\hat{x}_{D1}-g_{D6}\hat{x}%
_{D2}-\hat{x}_{D5})=-1.57\pm 0.09$ dB, \ \ $V(\hat{p}_{D5}-\hat{x}%
_{D4}-g_{D6}\hat{x}_{D7}-g_{D6}\hat{x}_{D8})=-1.53\pm 0.09$ dB. From
these measured results we can calculated the combinations of the
correlation variances in the left-hand sides of the inequalities
(3a)-(3g) and
(4a)-(4i), which are $0.68\pm 0.02$, $0.83\pm 0.02$, $0.82\pm 0.02$, $%
0.81\pm 0.02$, $0.82\pm 0.02$, $0.87\pm 0.02$, $0.75\pm 0.02$, for
the linear cluster and $0.84\pm 0.02$, $0.85\pm 0.02$, $0.96\pm
0.02$, $0.97\pm
0.02$, $0.95\pm 0.02$, $0.96\pm 0.02$, $0.96\pm 0.02$, $0.83\pm 0.02$, $%
0.83\pm 0.02$ for the two-diamond cluster, respectively. All these
values are smaller than the boundary. It means that the prepared two
types of CV cluster states satisfy the inseparability criteria for
verifying multipartite CV entanglement, so the spatially separated
eight-partite entangled states of photonic qumodes are
experimentally obtained. In the
experiment we detected the correlation variances under $g_{L1-L8}=1$, $%
g_{D1-D5}=1$\ and $g_{D6}=g_{D6}^{opt}=0.60\pm 0.02$. For our
system, the total transmission efficiency of squeezed beams are
about $87\%$\ and the detection efficiency is about $90\%$, which
lead to the efficient squeezing parameter is $r_{e}=0.30$ which is
smaller than the initially measured squeezing parameter. When the
gain factors except $g_{D6}$\ are taken as 1 and only
$g_{D6}^{opt}$\ is utilized, all inequalities in equation (3) and
equation (4) are satisfied. If $r_{e}>0.35$, the unite gain factor of $%
g_{D6}=1$\ can be chosen also (see Supplementary Material).

In the conclusion, we have experimentally prepared two types of
spatially separated eight-partite CV cluster entangled states for
photonic qumodes by using eight quadrature squeezed states of light
and a specifically designed optical beam-splitter network. The
multipartite entangled states are the essential resources to
construct a variety of CVQI networks. So far, the single-mode
squeezed states over 12.7 dB \cite{Eberle2010} and the two-mode
squeezed states over 8.1 dB \cite{Yan2012} have been experimentally
generated, respectively, based on which and using the presented
scheme the CV cluster states with more space-separable qumodes and
higher entanglement can be obtained. The complexity and versatility
of CV multipartite entanglement for photonic qumodes not only offer
richly potential applications in QC and QI, but also provide the
basic and handleable entangled quantum states which can be an
important tool for further studying the amazing and attractive
quantum entanglement phenomena.

This research was supported by the National Basic Research Program
of China (Grant No. 2010CB923103) and NSFC (Grant Nos. 11174188,
61121064).

\newpage

\centerline{\Large Supplementary Information for ``Experimental
preparation of eight-partite}
\centerline{\Large  linear and
two-diamond shape cluster states for photonic qumodes"}

\begin{center}
Xiaolong Su, Yaping Zhao, Shuhong Hao, Xiaojun Jia, Changde Xie, and Kunchi
Peng*\\[0pt]
State Key Laboratory of Quantum Optics and Quantum Optics Devices, Institute
of Opto-Electronics, Shanxi University, Taiyuan, 030006, People's Republic
of China\\[0pt]
*e-mail:kcpeng@sxu.edu.cn
\end{center}

\section{Unitary matrix for generating CV eight-partite cluster states}

According to the proposal of Peter van Loock et al \cite{Loock2007},
CV cluster states of photonic qumodes can be created via a general
linear-optics transformation of $\hat{p}$-squeezed input modes. If $\hat{a}%
_{l}=e^{+r}\hat{x}_{l}^{(0)}+ie^{-r}\hat{p}_{l}^{(0)}$ and a unitary matrix $%
U$ stand for the annihilation operator of the input modes and the
linear-optical transformation respectively, the output modes after
the
transformation are expressed by $\hat{b}_{k}=\sum\nolimits_{l}U_{kl}\hat{a}%
_{l}$, which are the CV cluster states \cite{Loock2007,Zhang2006}.
The CV cluster states satisfy
$I$Im[$UB_{in}$]$-A$Re[$UB_{in}$]$\rightarrow 0$ in
the limit of infinite squeezing, where $I$ is the identity matrix, $B_{in}=(%
\begin{array}{cccc}
\hat{a}_{1}, & \hat{a}_{2}, & \cdots , & \hat{a}_{n}%
\end{array}%
)^{T}$ is the matrix of input states, $A$ is the adjacency matrix \cite%
{Menicucci}. So we have $I$Im$U=A$Re$U,$ and the unitary matrix is
obtained

\begin{equation}
U=(I+iA)\text{Re}U.
\end{equation}

Based on the unitarity of matrix $U$, $UU^{\dagger }=I$, we have

\begin{equation}
\text{Re}U(\text{Re}U)^{T}=(I+A^{2})^{-1}.
\end{equation}%
In this case, we can obtain Re$U$ and $U$ from the adjacency matrix
$A$.

For n-partite cluster state, assuming

\begin{equation}
\text{Re}U=%
\begin{pmatrix}
\vec{\alpha}_{1}^{T} \\
\vec{\alpha}_{2}^{T} \\
\vdots \\
\vec{\alpha}_{n}^{T}%
\end{pmatrix}%
\end{equation}%
where $\vec{\alpha}_{i}^{T}=(%
\begin{array}{cccc}
\alpha _{i1}, & \alpha _{i2}, & \cdots , & \alpha _{in}%
\end{array}%
)$ is a real vectors. According to Eq. (2), we have $\vec{\alpha}_{i}^{T}%
\vec{\alpha}_{j}=(I+A^{2})_{ij}^{-1}$ ($i,j=1,\cdots ,n$), where the
numbers of these equations are $n(n+1)/2$ according to the symmetry
of matrix. Since there are $n^{2}$ unknown numbers in all these
equations, we need $n(n-1)/2$ conditions to solve the equations. For
simplicity and without lossing generality, some unknown numbers in
the equations are chosen to be 0 when we solve the equations.

For CV eight-partite linear cluster state, the adjacency matrix can
be written as

\begin{equation}
A=\ \left(
\begin{array}{cccccccc}
0 & 1 & 0 & 0 & 0 & 0 & 0 & 0 \\
1 & 0 & 1 & 0 & 0 & 0 & 0 & 0 \\
0 & 1 & 0 & 1 & 0 & 0 & 0 & 0 \\
0 & 0 & 1 & 0 & 1 & 0 & 0 & 0 \\
0 & 0 & 0 & 1 & 0 & 1 & 0 & 0 \\
0 & 0 & 0 & 0 & 1 & 0 & 1 & 0 \\
0 & 0 & 0 & 0 & 0 & 1 & 0 & 1 \\
0 & 0 & 0 & 0 & 0 & 0 & 1 & 0%
\end{array}%
\right)
\end{equation}%
we have

\begin{equation}
(I+A^{2})^{-1}=\left(
\begin{array}{cccccccc}
\frac{21}{34} & 0 & -\frac{4}{17} & 0 & \frac{3}{34} & 0 &
\frac{-1}{34} & 0
\\
0 & \frac{13}{34} & 0 & \frac{-5}{34} & 0 & \frac{1}{17} & 0 &
\frac{-1}{34}
\\
-\frac{4}{17} & 0 & \frac{8}{17} & 0 & -\frac{3}{17} & 0 &
\frac{1}{17} & 0
\\
0 & \frac{-5}{34} & 0 & \frac{15}{34} & 0 & -\frac{3}{17} & 0 &
\frac{3}{34}
\\
\frac{3}{34} & 0 & -\frac{3}{17} & 0 & \frac{15}{34} & 0 &
\frac{-5}{34} & 0
\\
0 & \frac{1}{17} & 0 & -\frac{3}{17} & 0 & \frac{8}{17} & 0 &
-\frac{4}{17}
\\
\frac{-1}{34} & 0 & \frac{1}{17} & 0 & \frac{-5}{34} & 0 &
\frac{13}{34} & 0
\\
0 & \frac{-1}{34} & 0 & \frac{3}{34} & 0 & -\frac{4}{17} & 0 & \frac{21}{34}%
\end{array}%
\right)
\end{equation}

The matrix elements in Eq. (3) are obtained by the following way.
Considering the symmetry, we start from the middle row, $\vec{\alpha}%
_{4}^{T}=(%
\begin{array}{cccccccc}
\alpha _{41}, & \alpha _{42}, & \alpha _{43}, & \alpha _{44}, &
\alpha _{45},
& \alpha _{46}, & \alpha _{47}, & \alpha _{48}%
\end{array}%
)$. We apply seven initial conditions, $\alpha _{41}=\alpha
_{42}=\alpha _{43}=\alpha _{44}=\alpha _{46}=\alpha _{47}=\alpha
_{48}=0$, then it
becomes $\vec{\alpha}_{4}^{T}=(%
\begin{array}{cccccccc}
0 & 0 & 0 & 0 & \alpha _{45} & 0 & 0 & 0%
\end{array}%
).$ According to equations $\vec{\alpha}_{4}^{T}\vec{\alpha}%
_{4}=(I+A^{2})_{44}^{-1}=\frac{15}{34}$, first unknown number $\alpha _{45}=-%
\sqrt{\frac{15}{34}}$ is obtained. Then we apply six conditions,
$\alpha _{51}=\alpha _{52}=\alpha _{53}=\alpha _{56}=\alpha
_{57}=\alpha _{58}=0$ on
$\vec{\alpha}_{5}^{T}$, we have $\vec{\alpha}_{5}^{T}=(%
\begin{array}{cccccccc}
0 & 0 & 0 & \alpha _{54} & \alpha _{55} & 0 & 0 & 0%
\end{array}%
)$. Using equations $\vec{\alpha}_{5}^{T}\vec{\alpha}%
_{5}=(I+A^{2})_{55}^{-1}=\frac{15}{34}$ and $\alpha _{5}^{T}\alpha
_{4}=(I+A^{2})_{54}^{-1}=0$, two unknown numbers $\alpha _{55}=0$,
$\alpha _{54}=-\sqrt{\frac{15}{34}}$ are gotten, and so on.

After all the unknown numbers in Eq. (3) are obtained, the unitary
matrix is given from Eq. (1). The unitary matrix for eight-partite
linear cluster state with eight $\hat{p}$-squeezed states to be the
input states are expressed

\begin{equation}
U_{L}^{p}=\left(
\begin{array}{cccccccc}
\frac{1}{\sqrt{2}} & \frac{i}{\sqrt{3}} & \frac{1}{\sqrt{10}} & -\sqrt{\frac{%
3}{170}} & i\sqrt{\frac{5}{102}} & 0 & 0 & 0 \\
\frac{i}{\sqrt{2}} & \frac{1}{\sqrt{3}} & \frac{-i}{\sqrt{10}} & i\sqrt{%
\frac{3}{170}} & \sqrt{\frac{5}{102}} & 0 & 0 & 0 \\
0 & \frac{i}{\sqrt{3}} & -\sqrt{\frac{2}{5}} & \sqrt{\frac{6}{85}} & -i\sqrt{%
\frac{10}{51}} & 0 & 0 & 0 \\
0 & 0 & -i\sqrt{\frac{2}{5}} & -3i\sqrt{\frac{3}{170}} & -\sqrt{\frac{15}{34}%
} & 0 & 0 & 0 \\
0 & 0 & 0 & -\sqrt{\frac{15}{34}} & -3i\sqrt{\frac{3}{170}} & i\sqrt{\frac{2%
}{5}} & 0 & 0 \\
0 & 0 & 0 & -i\sqrt{\frac{10}{51}} & \sqrt{\frac{6}{85}} &
\sqrt{\frac{2}{5}}
& \frac{i}{\sqrt{3}} & 0 \\
0 & 0 & 0 & \sqrt{\frac{5}{102}} & i\sqrt{\frac{3}{170}} & \frac{i}{\sqrt{10}%
} & \frac{1}{\sqrt{3}} & \frac{-i}{\sqrt{2}} \\
0 & 0 & 0 & i\sqrt{\frac{5}{102}} & -\sqrt{\frac{3}{170}} & \frac{-1}{\sqrt{%
10}} & \frac{i}{\sqrt{3}} & \frac{-1}{\sqrt{2}}%
\end{array}%
\right)
\end{equation}

In the experiment, we prepared four $\hat{x}$-squeezed states, $\hat{a}%
_{m}=e^{-r}\hat{x}_{m}^{(0)}+ie^{+r}\hat{p}_{m}^{(0)}$
$(m=1,3,5,7)$, and
four $\hat{p}$-squeezed states, $\hat{a}_{n}=e^{+r}\hat{x}_{n}^{(0)}+ie^{-r}%
\hat{p}_{n}^{(0)}(n=2,4,6,8)$ with four NOPAs, respectively. The
transformation between $\hat{x}$-squeezed state and
$\hat{p}$-squeezed state can be achieved via a Fourier
transformation. Applying Fourier
transformation on modes $\hat{a}_{1},\hat{a}_{3},\hat{a}_{5}$ and $\hat{a}%
_{7}$, which corresponds to multiplying $i$ to the values of columns
1, 3, 5 and 7 in the unitary matrix $U_{L}^{p}$, we obtain the
unitary matrix of CV eight-partite linear cluster state for our
experimental system, which is

\begin{equation}
U_{L}=\left(
\begin{array}{cccccccc}
\frac{i}{\sqrt{2}} & \frac{i}{\sqrt{3}} & \frac{i}{\sqrt{10}} & \sqrt{\frac{3%
}{170}} & \sqrt{\frac{5}{102}} & 0 & 0 & 0 \\
\frac{-1}{\sqrt{2}} & \frac{1}{\sqrt{3}} & \frac{1}{\sqrt{10}} & -i\sqrt{%
\frac{3}{170}} & -i\sqrt{\frac{5}{102}} & 0 & 0 & 0 \\
0 & \frac{i}{\sqrt{3}} & -i\sqrt{\frac{2}{5}} & -\sqrt{\frac{6}{85}} & -%
\sqrt{\frac{10}{51}} & 0 & 0 & 0 \\
0 & 0 & \sqrt{\frac{2}{5}} & 3i\sqrt{\frac{3}{170}} &
i\sqrt{\frac{15}{34}}
& 0 & 0 & 0 \\
0 & 0 & 0 & \sqrt{\frac{15}{34}} & -3\sqrt{\frac{3}{170}} & i\sqrt{\frac{2}{5%
}} & 0 & 0 \\
0 & 0 & 0 & i\sqrt{\frac{10}{51}} & -i\sqrt{\frac{6}{85}} & \sqrt{\frac{2}{5}%
} & \frac{1}{\sqrt{3}} & 0 \\
0 & 0 & 0 & -\sqrt{\frac{5}{102}} & \sqrt{\frac{3}{170}} & \frac{i}{\sqrt{10}%
} & \frac{-i}{\sqrt{3}} & \frac{-i}{\sqrt{2}} \\
0 & 0 & 0 & -i\sqrt{\frac{5}{102}} & i\sqrt{\frac{3}{170}} & \frac{-1}{\sqrt{%
10}} & \frac{1}{\sqrt{3}} & \frac{-1}{\sqrt{2}}%
\end{array}%
\right)
\end{equation}

Similarly, the unitary matrix for generating CV eight-partite
two-diamond shape cluster state can be obtained with the same way.
The calculated
results show that $U_{D}$ equals to $U_{F}U_{L}$, where $U_{F}=diag%
\{-1,-i,i,1,1,i,-i,-1\}$. The unitary matrix $U_{D}$ is given by

\begin{equation}
U_{D}=\left(
\begin{array}{cccccccc}
\frac{-i}{\sqrt{2}} & \frac{-i}{\sqrt{3}} & \frac{-i}{\sqrt{10}} & -\sqrt{%
\frac{3}{170}} & -\sqrt{\frac{5}{102}} & 0 & 0 & 0 \\
\frac{i}{\sqrt{2}} & \frac{-i}{\sqrt{3}} & \frac{-i}{\sqrt{10}} & -\sqrt{%
\frac{3}{170}} & -\sqrt{\frac{5}{102}} & 0 & 0 & 0 \\
0 & \frac{-1}{\sqrt{3}} & \sqrt{\frac{2}{5}} & -i\sqrt{\frac{6}{85}} & -i%
\sqrt{\frac{10}{51}} & 0 & 0 & 0 \\
0 & 0 & \sqrt{\frac{2}{5}} & 3i\sqrt{\frac{3}{170}} &
i\sqrt{\frac{15}{34}}
& 0 & 0 & 0 \\
0 & 0 & 0 & \sqrt{\frac{15}{34}} & -3\sqrt{\frac{3}{170}} & i\sqrt{\frac{2}{5%
}} & 0 & 0 \\
0 & 0 & 0 & -\sqrt{\frac{10}{51}} & \sqrt{\frac{6}{85}} &
i\sqrt{\frac{2}{5}}
& \frac{i}{\sqrt{3}} & 0 \\
0 & 0 & 0 & i\sqrt{\frac{5}{102}} & -i\sqrt{\frac{3}{170}} & \frac{1}{\sqrt{%
10}} & \frac{-1}{\sqrt{3}} & \frac{-1}{\sqrt{2}} \\
0 & 0 & 0 & i\sqrt{\frac{5}{102}} & -i\sqrt{\frac{3}{170}} & \frac{1}{\sqrt{%
10}} & \frac{-1}{\sqrt{3}} & \frac{1}{\sqrt{2}}%
\end{array}%
\right)
\end{equation}

\section{Inseparability criteria}

According to the inseparability criteria for CV multipartite
entangled states proposed by van Loock and Furusawa
\cite{Loock2003}, we deduced the concrete inseparability conditions
for CV eight-partite linear cluster state, which are Eqs. (3) in the
main text. For any separable quantum state,
its total density operator can be written as $\hat{\rho}=\sum_{i}\eta _{i}%
\hat{\rho}_{i,k,...,m}\otimes \hat{\rho}_{i,l,...,n}$ with a
distinct pair of ``separable modes" ($m,n$) and the other modes
$k\neq l$ [see equation (25) in Ref. 2.], where $\eta _{i}$
represent the mixture of these separable
states. For any combinations $\hat{u}=h_{1}\hat{x}_{1}+h_{2}\hat{x}%
_{2}+\cdots +h_{N}\hat{x}_{N}$ and $\hat{v}=g_{1}\hat{p}_{1}+g_{2}\hat{p}%
_{2}+\cdots +g_{N}\hat{p}_{N}$, the inseparability criteria are
expressed by \cite{Loock2003}

\begin{equation}
V(\hat{u})+V(\hat{v})<\frac{1}{2}(\left\vert
h_{m}g_{m}+\sum\nolimits_{k}h_{k}g_{k}\right\vert +\left\vert
h_{n}g_{n}+\sum\nolimits_{l}h_{l}g_{l}\right\vert )
\end{equation}%
Based on this equation, we deduced the inseparability criteria for
CV eight-partite linear and two-diamond shape cluster states, which
are Eqs. (3) and (4) in the main text, respectively.

The violation of each inequality in Eqs. (3) can be used to verify
variously partially separable states, that are

\begin{eqnarray}
\text{violation of }(3a) &\Longrightarrow
&\hat{\rho}=\sum\nolimits_{i}\eta
_{i}\hat{\rho}_{i,k,...,1}\otimes \hat{\rho}_{i,l,...,2} \\
\text{violation of }(3b) &\Longrightarrow
&\hat{\rho}=\sum\nolimits_{i}\eta
_{i}\hat{\rho}_{i,k,...,2}\otimes \hat{\rho}_{i,l,...,3}  \notag \\
\text{violation of }(3c) &\Longrightarrow
&\hat{\rho}=\sum\nolimits_{i}\eta
_{i}\hat{\rho}_{i,k,...,3}\otimes \hat{\rho}_{i,l,...,4}  \notag \\
\text{violation of }(3d) &\Longrightarrow
&\hat{\rho}=\sum\nolimits_{i}\eta
_{i}\hat{\rho}_{i,k,...,4}\otimes \hat{\rho}_{i,l,...,5}  \notag \\
\text{violation of }(3e) &\Longrightarrow
&\hat{\rho}=\sum\nolimits_{i}\eta
_{i}\hat{\rho}_{i,k,...,5}\otimes \hat{\rho}_{i,l,...,6}  \notag \\
\text{violation of }(3f) &\Longrightarrow
&\hat{\rho}=\sum\nolimits_{i}\eta
_{i}\hat{\rho}_{i,k,...,6}\otimes \hat{\rho}_{i,l,...,7}  \notag \\
\text{violation of }(3g) &\Longrightarrow
&\hat{\rho}=\sum\nolimits_{i}\eta _{i}\hat{\rho}_{i,k,...,7}\otimes
\hat{\rho}_{i,l,...,8}  \notag
\end{eqnarray}%
If all inequalities in Eqs. (3) are satisfied, we can confirm that
all eight states are inseparable and thus they construct a CV
eight-partite linear cluster entangled state.

We also deduced the inseparability criteria for CV eight-partite
two-diamond shape linear cluster state, which are Eqs. (4) in the
main text. Similarly, if the inequalities are violated the partially
separable states are expressed by

\begin{eqnarray}
\text{violation of }(4a) &\Longrightarrow
&\hat{\rho}=\sum\nolimits_{i}\eta
_{i}\hat{\rho}_{i,k,...,1}\otimes \hat{\rho}_{i,l,...,3} \\
\text{violation of }(4b) &\Longrightarrow
&\hat{\rho}=\sum\nolimits_{i}\eta
_{i}\hat{\rho}_{i,k,...,2}\otimes \hat{\rho}_{i,l,...,3}  \notag \\
\text{violation of }(4c) &\Longrightarrow
&\hat{\rho}=\sum\nolimits_{i}\eta
_{i}\hat{\rho}_{i,k,...,1}\otimes \hat{\rho}_{i,l,...,4}  \notag \\
\text{violation of }(4d) &\Longrightarrow
&\hat{\rho}=\sum\nolimits_{i}\eta
_{i}\hat{\rho}_{i,k,...,2}\otimes \hat{\rho}_{i,l,...,4}  \notag \\
\text{violation of }(4e) &\Longrightarrow
&\hat{\rho}=\sum\nolimits_{i}\eta
_{i}\hat{\rho}_{i,k,...,4}\otimes \hat{\rho}_{i,l,...,5}  \notag \\
\text{violation of }(4f) &\Longrightarrow
&\hat{\rho}=\sum\nolimits_{i}\eta
_{i}\hat{\rho}_{i,k,...,5}\otimes \hat{\rho}_{i,l,...,7}  \notag \\
\text{violation of }(4g) &\Longrightarrow
&\hat{\rho}=\sum\nolimits_{i}\eta
_{i}\hat{\rho}_{i,k,...,5}\otimes \hat{\rho}_{i,l,...,8}  \notag \\
\text{violation of }(4h) &\Longrightarrow
&\hat{\rho}=\sum\nolimits_{i}\eta
_{i}\hat{\rho}_{i,k,...,6}\otimes \hat{\rho}_{i,l,...,7}  \notag \\
\text{violation of }(4i) &\Longrightarrow
&\hat{\rho}=\sum\nolimits_{i}\eta _{i}\hat{\rho}_{i,k,...,6}\otimes
\hat{\rho}_{i,l,...,8}  \notag
\end{eqnarray}%
When all inequalities in Eqs. (4) are satisfied CV eight-partite
two-diamond shape cluster entanglement is demonstrated.

\begin{figure}[tbp]
\begin{center}
\includegraphics[width=15cm,height=10cm]{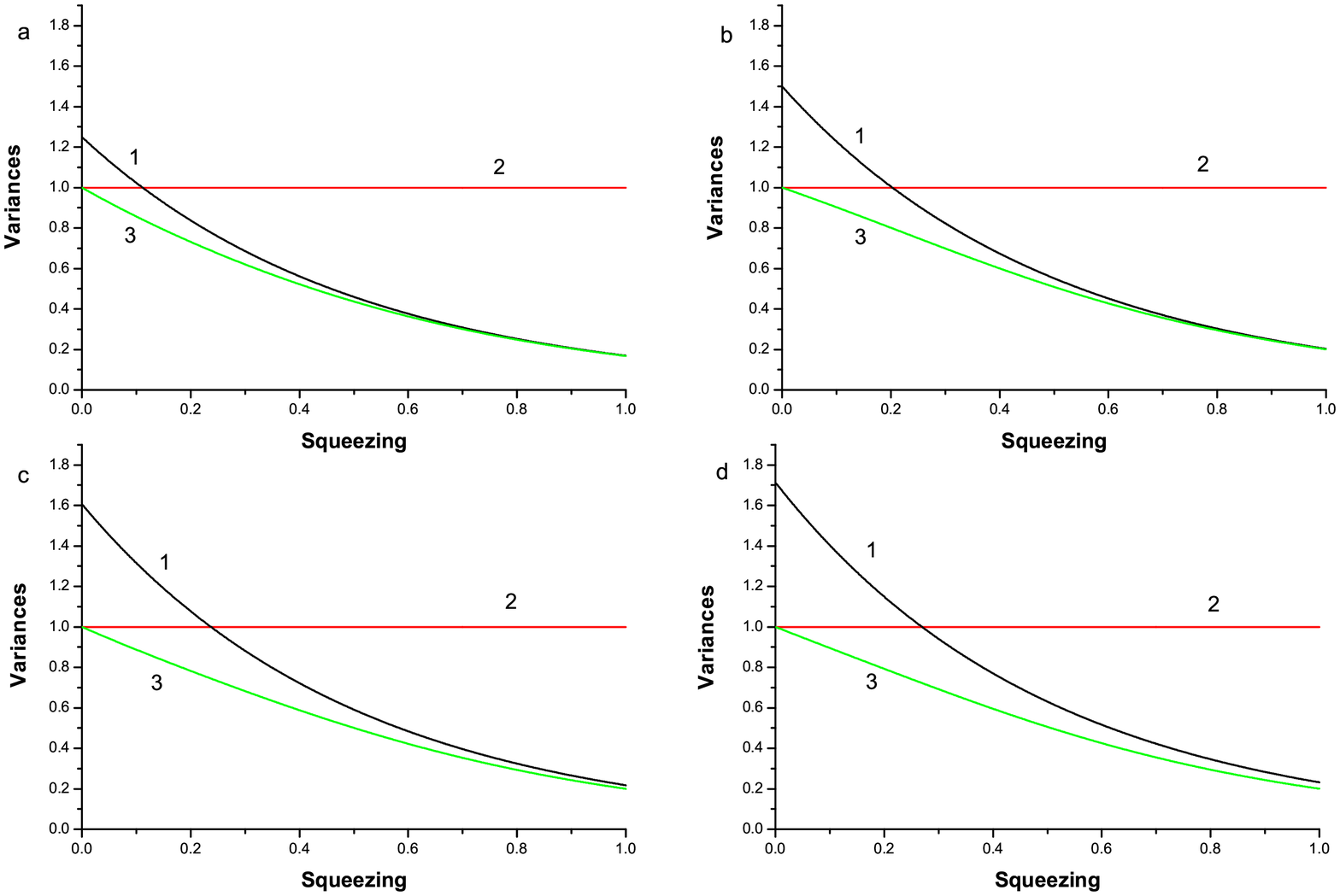}
\end{center}
\caption{The dependence of eight-partite linear cluster state to
squeezing parameter, a-d are corresponding to inequalities
(3a)-(3d), respectively. Lines 1 and 3 are left-hand sides of
inequalities with unit gain and optimal gain, respectively, line 2
are right-hand sides of inequalities.}
\end{figure}

Calculating the minimal values of the left-hand sides of the
inequalities versus the gain factors ($g_{L1-L8}$\ for the linear
shape and $g_{D1-D6}$
for the two-diamond shape) we can obtain the optimal gain factors $%
g_{L1-L8}^{opt}$ and $g_{D1-D6}^{opt}$ for achieving the detections
of the minimal correlation variances. For the linear
($g_{L1-L8}^{opt}$ )\ and the
two diamond ($g_{D1-D6}^{opt}$) cluster states the optimal gain factors are $%
g_{L1}=g_{L8}=\frac{21(e^{4r}-1)}{13+21e^{4r}}$, $g_{L2}=g_{L7}=\frac{%
13(e^{4r}-1)}{21+13e^{4r}}$, $g_{L3}=g_{L6}=\frac{8(e^{4r}-1)}{9+8e^{4r}}$, $%
g_{L4}=g_{L5}=\frac{15(e^{4r}-1)}{19+15e^{4r}}$\ and $g_{D1}=\frac{%
15(e^{4r}-1)}{19+15e^{4r}}$, $g_{D2}=\frac{21(e^{4r}-1)}{13+21e^{4r}}$, $%
g_{D3}=\frac{9(e^{4r}-1)}{8+9e^{4r}}$, $g_{D4}=\frac{9(e^{8r}-1)}{%
7+18e^{4r}+9e^{8r}}$,
$g_{D5}=\frac{3(3e^{8r}-2e^{4r}-1)}{7+18e^{4r}+9e^{8r}} $,
$g_{D6}=\frac{4(e^{4r}-1)}{13+4e^{4r}}$, respectively. The
dependence of inseparability criteria of CV eight-partite linear
cluster state on the squeezing parameter is shown in figure 1. In
Fig. 1, (a), (b), (c) and (d) correspond to inequalities (3a), (3b),
(3c) and (3d), respectively. Lines 1 and 3 are the variance
combinations in the left-hand sides of inequalities with unit gain
and the optimal gain, respectively. Line 2 stands for the boundary
corresponding to right-hand sides of inequalities. Since the other
inequalities have same variances with (3a)-(3c), we only show the
four results here. For inequalities (3a)-(3d), if the unit gain
factor is used only when $r>0.11$, $r>0.20$, $r>0.24$, and $r>0.27$
these inequalities are satisfied. However, if the optimal gain
factors are chosen, these criterion inequalities are always
satisfied for any $r>0,$ i.e. the presented system can prepare the
CV eight-partite cluster state without the lower limitation of the
squeezing degree for the input squeezed state.

\begin{figure}[tbp]
\begin{center}
\includegraphics[width=15cm,height=10cm]{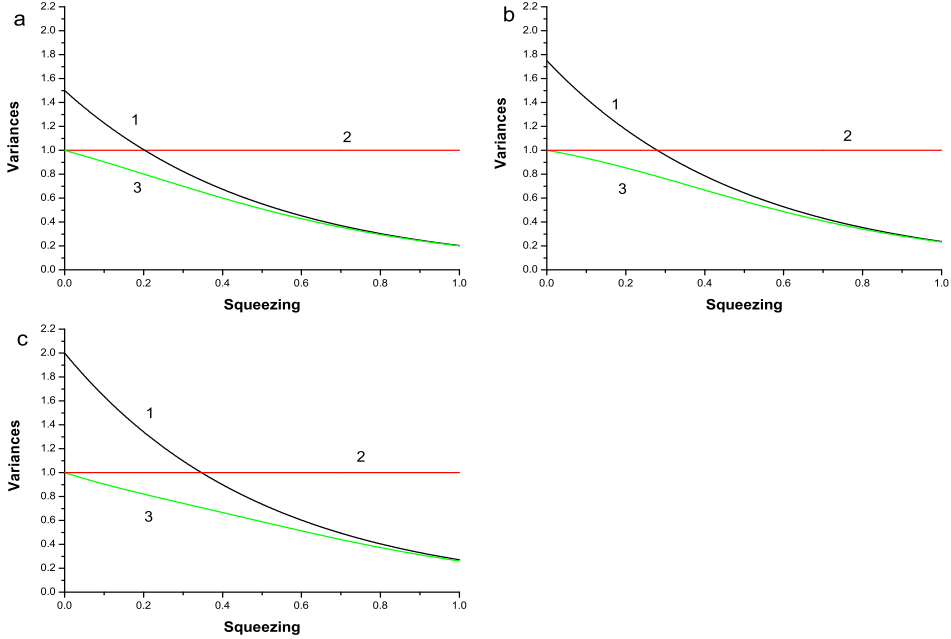}
\end{center}
\caption{The dependence of eight-partite two-diamond shape cluster
state to squeezing parameter, a, b and c are corresponding to
inequalities (4a), (4c), and (4e), respectively. Lines 1 and 3 are
left-hand sides of inequalities with unit gain and optimal gain,
respectively, line 2 are right-hand sides of inequalities.}
\end{figure}
The dependence of inseparability criteria of CV eight-partite
two-diamond shape cluster state to the squeezing parameter is shown
in figure 2. Since the variances of inequalities (4a), (4b), (4h)
and (4i) are the same, and that of (4c), (4d), (4f) and (4g) are
also the same, we only plot the correlation variances of the
inequalities (4a), (4c) and (4e) in Fig. 2a, 2b and 2c,
respectively. From Fig. 2a, 2b and 2c, we can see that when the
squeezing parameters $r>0.20$ for Fig. 2a, $r>0.28$ for Fig. 2b, and
$r>0.35$ for Fig. 2c, the correlation variances (curve 1) are
smaller than the boundary for the case of $g=1$. It means that if
taking $g=1$, a lower limitation for the squeezing parameter is
required to satisfy each of the inseparability criteria. However, if
taking the optimal gain factor $g^{opt}$ (curve 3) all variances are
below the boundary for any values of $r>0$.

\section{Experimental details}

The four NOPAs were constructed in identical configuration, each of
which
consists of an $\alpha $-cut type-II KTP crystal and a concave mirror \cite%
{Wang20102}. The front face of the KTP was coated to be used for the
input coupler and the concave mirror serves as the output coupler of
the squeezed states, which was mounted on a piezo-electric
transducer for locking
actively the cavity length of NOPA on resonance with the injected signal at $%
1080$ nm. The transmissions of the input coupler for the pump laser
at 540 nm and the injected signal at 1080 nm are $99.8\%$ and
$0.04\%$, respectively. The transmissions of the output coupler at
540 nm and 1080 nm are $0.5\%$ and $5.2\%$, respectively. The
finesses of the NOPA for 540 nm and 1080 nm are $3$ and $117$,
respectively. Through the intracavity parametric down conversion
process of type-II phase match, the two quadrature squeezed states
of light at $1080$ nm were produced \cite{Reid, Yun2000}. The
squeezing parameter $r$\ depends on the strength and the time of
parametric interaction in the NOPA. In the calculations we have
assumed that squeezing parameter for the eight squeezed states is
identical for simplicity and without lossing the generality. The
requirement is easy to be reached in the experiments if the four
NOPAs were constructed in identical configuration and the
intracavity losses of the signal and the idler modes in each NOPA
were balanced. The experimentally measured squeezing degrees of the
output fields from four NOPAs are about $4.30\pm 0.07$ dB below the
corresponding QNL, which are shown in Fig. 3, a and b stand for
quadrature-amplitude and quadrature-phase squeezing, respectively.

\begin{figure}[tbp]
\begin{center}
\includegraphics[width=10cm,height=6cm]{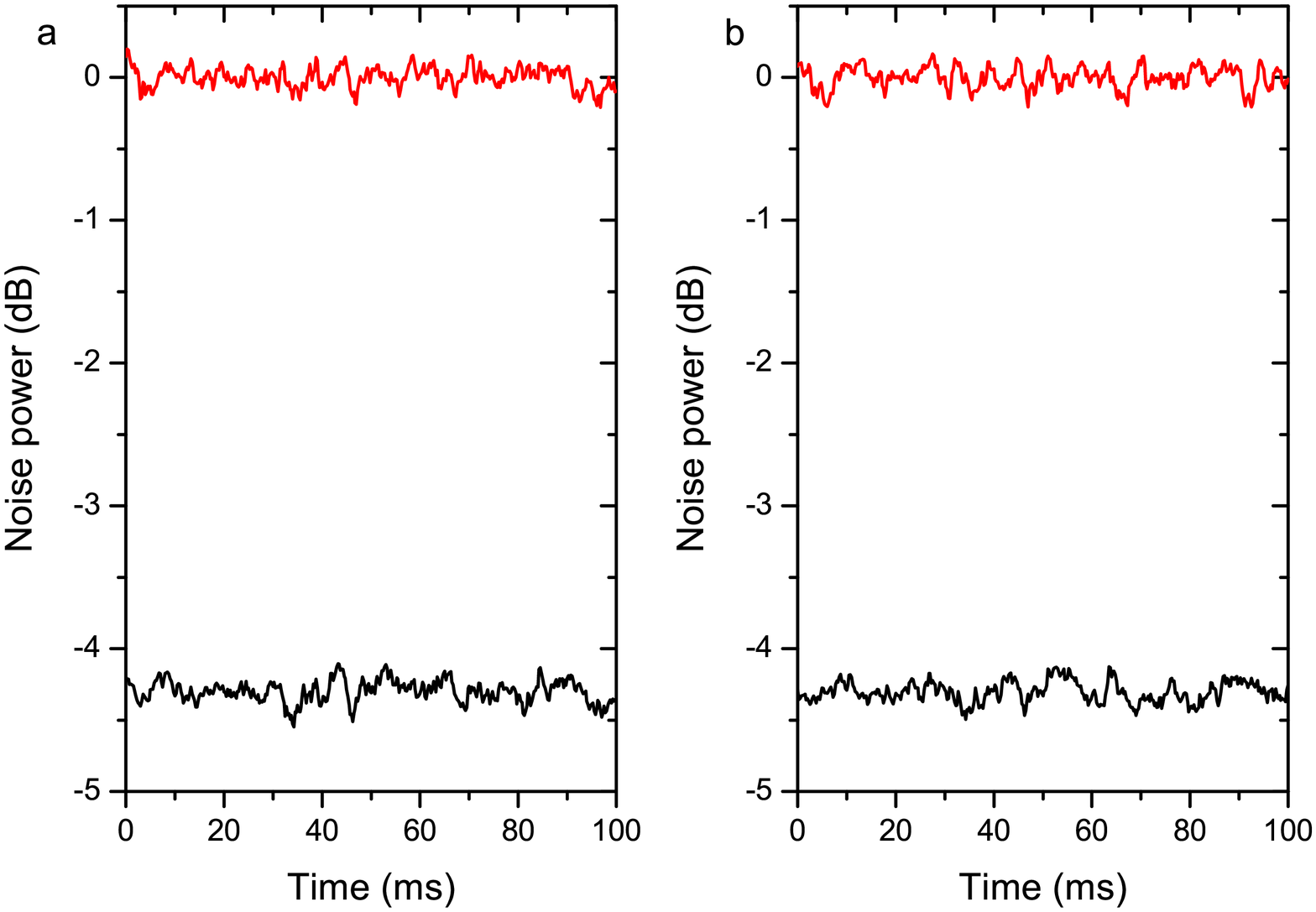}
\end{center}
\caption{The measured noise power of initial squeezed state. a and b
are amplitude and phase quadrature, respectively. Red and black
lines are the noise power for the QNL and the squeezed state,
respectively.}
\end{figure}

Four NOPAs are locked individually by using Pound-Drever-Hall (PDH)
method with a phase modulation of 56 MHz on 1080 nm laser beam
\cite{Pound}. Each NOPA is operated at deamplification condition,
which corresponds to lock the relative phase between the pump laser
and the injected signal to (2n+1)$\pi $ (n is the integer). In the
experiment, the relative phase locking is completed with a lock-in
amplifier, where a signal at 15 kHz is modulated on the pump light
by the piezo-electric transducer (PZT) mounted on a reflection
mirror which is placed in the optical path of the pump laser and
then the error signal is fed back to the other PZT which is mounted
on a mirror placed in the optical path of the injected beam.

In the beam-splitter network used to prepare eight-partite cluster
states, T1 is phase-locked to zero, T2-T7 is phase-locked to $\pi
/2$. The zero phase difference (T1) between two interfered beams on
a beam-splitter is locked by a lock-in amplifier with a modulation
of 12 kHz. The $\pi /2$ phase difference (T2-T7) is locked by DC
locking technique, where the photocurrent signal of the interference
fringe is fed back to the PZT mounted on a mirror which is placed
before the beam-splitter.

In the homodyne detection system, zero phase difference for the
measurement of quadrature-amplitude is locked by PDH technique with
a phase modulation of 10.9 MHz on local oscillator beam. The $\pi
/2$ phase for the measurement of quadrature-phase is locked by using
DC locking technique too.

\end{document}